\documentclass[sigconf,nonacm]{acmart}
\AtBeginDocument{%
  }

\setcopyright{acmlicensed}
\copyrightyear{2018}
\acmYear{2018}
\acmDOI{XXXXXXX.XXXXXXX}
\acmConference[Conference acronym 'XX]{Make sure to enter the correct
  conference title from your rights confirmation email}{June 03--05,
  2018}{Woodstock, NY}
\acmISBN{978-1-4503-XXXX-X/2018/06}

\usepackage{tabularx}
\usepackage{xcolor}
\usepackage{listings}
\providecolor{lightblue}{RGB}{235,245,255}

\usepackage{xcolor}

\definecolor{convgreen}{HTML}{48C92F}
\definecolor{dmorange}{HTML}{FFAA0A}
\definecolor{dragpurple}{HTML}{7040E8}

\newcommand{\fignum}[2]{\textcolor{#1}{\textbf{#2}}}

\begin{document}

\title{Say It or Grab It? Conversational Control of UI Adaptations in Mixed Reality}


\author{Sina Elahimanesh}
\email{sina.elahimanesh@cs.uni-saarland.de}
\affiliation{%
  \institution{Saarland Informatics Campus, Saarland University}
  \city{Saarbrücken}
  \country{Germany}
}

\author{João Marcelo Evangelista Belo}
\email{jbelo@cs.uni-saarland.de}
\affiliation{%
  \institution{Saarland Informatics Campus, Saarland University}
  \city{Saarbrücken}
  \country{Germany}
}

\author{Isidora Jeknić}
\email{jeknic@lst.uni-saarland.de}
\affiliation{%
  \institution{Department of Language Science and Technology, Saarland University}
  \city{Saarbrücken}
  \country{Germany}
}

\author{Alexander Koller}
\email{koller@coli.uni-saarland.de}
\affiliation{%
  \institution{Department of Language Science and Technology, Saarland University}
  \city{Saarbrücken}
  \country{Germany}
}

\author{Anna Maria Feit}
\email{feit@cs.uni-saarland.de}
\affiliation{%
  \institution{Saarland Informatics Campus, Saarland University}
  \city{Saarbrücken}
  \country{Germany}
}

\renewcommand{\shortauthors}{Elahimanesh et. al.}
\newcommand{\system}{ConverseMR}

\begin{abstract}
  Mixed reality interfaces must continuously adapt to a user's changing task and environment, yet adaptation policies are fixed in advance, either preset by the system or configured by the user. Natural language offers a way to state changing needs mid-task, but whether users can express interface behavior this way, and what it gains over direct manipulation, is unknown. We first elicited how users verbalize adaptation needs (N=19). We then built \system{}, which translates a spoken request into adaptation objectives and constraints and satisfies them continuously as the user's context changes. In a within-subjects study (N=24), conversational and direct control proved complementary: direct manipulation for precision, \system{} for reach and expressive breadth, with their combination rated highest. Spoken commands were rarely final and users frequently refined and replaced them upon experiencing the result, characteristic of, what we call, user-led adaptation, where users communicate adaption needs and the system identifies a solution.
\end{abstract}



\begin{teaserfigure}
  \includegraphics[width=\textwidth]{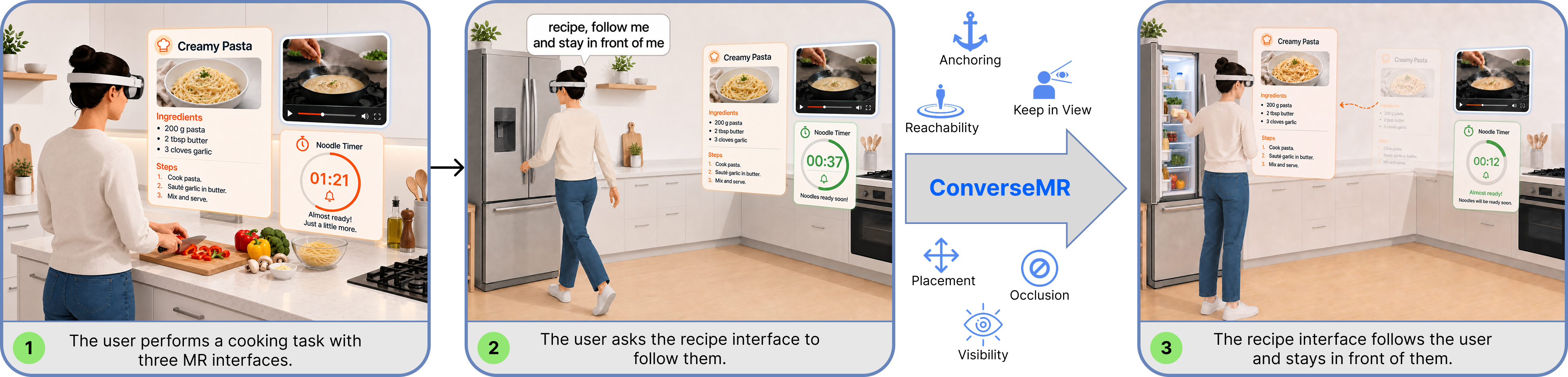}
  \caption{This paper explores the usefulness and usability of controlling the adaptive behavior of mixed reality interfaces through natural language. As an example, consider Alice who cuts vegetables, following an MR recipe (1). As she moves to the fridge to fetch more ingredients, she asks the recipe interface to follow her. Other UI elements have no relevance for this part of the task. We develop \system{} that translates such a natural language command into a set of adaptation objectives, constraints, and parameters. As a result, the recipe follows Alice, while the video and timer remain in its initial place (3). We use \system{} to empirically investigate the usefulness of conversational interaction for MR interface adaptation.}
  \label{fig:teaser}
\end{teaserfigure}


\maketitle


\section{Introduction}
\label{sec:introduction}

Mixed Reality (MR) interfaces need to be adaptive to be usable across changing contexts, tasks, individual user preferences or constraints~\cite{adaptive2022lindlbauer}. In recent years, research has proposed a range of approaches to adapt MR interfaces, in particular the spatial placement of UI elements: view management techniques that keep content visible and unoccluded~\cite{viewmanagement2001bell,viewmanagement_grasset2012,density_tatzgern2016}, world- and egocentric anchoring strategies\cite{track_rasch2025, hand_jannat2025, uimobility_pei2024, gazego_manakhov2024}, and optimization-based methods that place elements with respect to multiple objectives such as ergonomics, semantic relevance, or the user's task context~\cite{contextadapt_lindlbauer2019,semanticadapt_cheng2021,belo2022auit}. Typically, adaptation criteria (e.g. to avoid occlusion of task-relevant things, to follow the users movement, or to ensure semantic similarity, etc.) are pre-defined by researchers or designers. However, a user's context and personal preferences are rich and change continuously~\cite{belo2026userpreferencesuianchoring}, and determining which adaptation criteria apply to a specific user in a specific moment is error-prone. Recent work has therefore recognized the importance of letting end-users shape the spatial organization~\cite{multiagentic_li2026, paretoadapt_johns2023, song2025} and adaptive behavior~\cite{belo2026userpreferencesuianchoring} of MR interfaces themselves through direct manipulation and selecting from pre-configured configurations~\cite{belo2026userpreferencesuianchoring, hayatpur2019spatialmanipulation,uimobility_pei2024}.

To illustrate this, consider Alice cooking a meal, following an MR recipe with video instructions and a timer (\autoref{fig:teaser}). While she prepares ingredients at the counter, these elements are best anchored where they do not occlude her hands. When she walks to the fridge, she needs the ingredient list to come with her, while the video and timer are momentarily irrelevant. Detecting such a subtle shift in task, inferring the task-relevance of each element, and choosing a suitable adaptive behavior for each is difficult to do automatically in a domain-independent way without occasionally inferring the wrong intent. For decades, research on adaptive GUIs has shown that the accuracy and the predictability of an adaptation determine whether users benefit from it at all~\cite{gajos2006,gajos2008,findlater2008}, and that users may prefer an interface they control themselves over a more efficient one that adapts on their behalf~\cite{findlater2004}. In MR the stakes are higher. Adaptation is a precondition for usable interfaces rather than a convenience, since content has no fixed screen to occupy and must be placed continually with respect to a user moving through a changing physical world~\cite{adaptive2022lindlbauer}. A poor placement can occlude the real-world task the user is performing, or force them to interrupt it. What is needed is a way for users to state their changing needs without interrupting their activity, and a computational approach that turns those needs into interface adaptations.

Large language models (LLMs) have emerged as a promising way to translate intent expressed through natural language into system behavior and interface configurations~\cite{malleableInterfaces, multiagentic_li2026}, offering functional flexibility \cite{lehmann2024functional, lehmannDiss} and more broadly a new way to access and personalize computing~\cite{rost2025interactions} .
We therefore see natural-language interaction and LLMs as a promising technique for users to communicate changing interaction needs during MR use. Recent work has shown the feasibility of this for generating static MR
layouts~\cite{multiagentic_li2026}. Whether users can also control the adaptive
behavior of an interface this way, and how they express their interaction needs
when they do, is unclear though.

In this paper, we aim to understand whether natural-language based control of adaptive MR interfaces is usable and useful. Specifically, we want to understand whether users are able to verbalize their adaptation needs, and how
natural-language-based control compares to direct manipulation in terms of customization frequency, breadth of expressible adaptations, and usability.
We first conducted a formative study (N=19) in which participants performed an MR task under different mobility settings and were asked how they
would verbally instruct the interface to behave. 
A thematic analysis of 57 elicited responses yields a variety of adaptation needs, from which we derive five design requirements for conversational MR adaptation. 
Guided by these, we develop \system{}, a system that lets users control the adaptive behavior of MR interfaces through speech. \system{} uses a structured, few-shot prompt to extract a user's intent from a
spoken utterance into an intermediate representation of adaptation objectives,
constraints, and parameters, asking for clarification where the interpretation is uncertain and otherwise adapting the interface through multi-objective optimization based on AUIT~\cite{belo2022auit}. It outputs a summary of the adaption intent as feedback to the user, who can refine, replace, or revert their command, after experiencing the result.

In a within-subjects study with 24 participants, we let users engage in a representative multi-tasking scenario, systematically varying  interaction requirements and movement demands to elicit changing adaptation needs. We explore how participants use and rate \system compared to a direct manipulation interface, and a combination of both, to customize the adaptive behavior of three UI elements to the changes in their environment and task. . 
Interestingly,  Combined condition is rated highest throughout. Rather than one method dominating, participants identified trade-offs between them. Direct manipulation was preferred for precise, close-range
placement, for its determinism, and for its discoverability of functionalities.
\system{} on the other hand was preferred for its expressive breadth, for
leaving the hands free, and for reach without walking. 
manipulation wherever speaking aloud is undesirable. 
Despite these trade-offs, voice absorbed the majority of adaptation activity once it was available, and over half of our participants never used the graphical menu in the Combined condition.

We contribute (i) a characterization of how users verbalize MR
adaptation needs established through a formative study and extended through  observations during live interaction, (ii)
\system{}, a conversational system enabling intent-driven control of UI adaptation in mixed reality, and (iii) an empirical understanding of the trade-offs between conversational and direct control of interface adaptation, showing that they are complementary rather than competing.
More broadly, our work points to a shift in how adaptive interfaces might be
built. Rather than choosing between systems that adapt on the user's behalf and systems users configure by hand, \system{} lets users state their adaptation needs and leave it to the system to carry it out. This form of user-led adaptation is a promising direction for adaptive interfaces beyond MR.

\section{Related Work}



\subsection{Adaptive Mixed Reality Interfaces}
The adaptation of Mixed Reality interfaces has been an active topic of research since its early days, as changes in the user's context such as movement and physical constraints require UI modifications to continue to be usable \cite{adaptive2022lindlbauer}.
Pioneering work from Bell et al. focused on view management techniques to maintain virtual content in the user's field of view, prioritizing the avoidance of occlusion by updating the position, size, and transparency of the virtual content \cite{viewmanagement2001bell}.
Grasset et al. later extended this with an image-based method that analyses the environment and adapts UI positions in the 2D view plane of an AR browser~\cite{viewmanagement_grasset2012}, and Tatzgern et al. moved the approach into three dimensions~\cite{viewmanagement_tatzgern2014}.
Beyond position, researchers have adapted other UI properties: Tatzgern et al. adapted level of detail to information density~\cite{density_tatzgern2016}, and Lindlbauer et al. adapted both position and level of detail to the user's context using combinatorial optimization~\cite{contextadapt_lindlbauer2019}.
Subsequent work broadened the set of factors an MR interface can adapt to, including environment semantics~\cite{semanticadapt_cheng2021}, ergonomics~\cite{xrgonomics_belo2021, comfortable_cheng2022}, and the user's surroundings or social situation~\cite{situationadapt_li2024}.

Most of these systems formulate adaptation as an optimization problem.
Belo et al. consolidate these objectives in the AUIT toolkit, separating adaptation objectives, optimization method, and the mechanism by which the UI changes~\cite{belo2022auit}.
Johns et al. build on this with ParetoAdapt, which presents multiple adaptation proposals from which the user selects~\cite{paretoadapt_johns2023}.
More recently, symbolic optimization has been extended with LLM-based reasoning, for example to interpret the scene~\cite{situationadapt_li2024} or the user's preferences~\cite{multiagentic_li2026}.
Across this literature, the objectives that govern adaptation are authored by the designer at development time. The user's role is, at most, to select among system-generated proposals~\cite{paretoadapt_johns2023}; they cannot state what the interface should optimize for, nor revise it as their task changes. An exception is the recent work by ~\citet{multiagentic_li2026} which proposes a similar architecture to our \system{} system, translating natural language into a symbolic layout model and using the AUIT toolkit~\cite{belo2022auit} to solve and execute the resulting model in XR. Rather than optimizing static UI layouts, we focus on enabling users to (re-)specify the adaptive behavior of UI elements as their task and interaction needs change over time.

\subsection{Spatial Reference Frames for UI anchoring}
Early view-management techniques proposed UI positioning strategies to improve the usability of XR systems, but only recently have researchers investigated how the choice of frame of reference itself affects usability. The question is particularly relevant under locomotion, where content can be anchored to the physical environment, to the user's body, or to physical objects \cite{track_rasch2025, hand_jannat2025, uimobility_pei2024, gazego_manakhov2024, belo2026userpreferencesuianchoring}. \emph{World-anchored} interfaces hold a stable position relative to the environment, and \citet{gazego_manakhov2024} show that target acquisition becomes faster and more precise as a target's stability relative to the environment increases. \emph{Egocentric} frames, in contrast, move with the user: head-anchoring keeps content within the field of view for quick glances, torso-anchoring offers a more stable egocentric reference, and hand-anchoring supports on-demand access to small elements \cite{track_rasch2025, hand_jannat2025, belo2026userpreferencesuianchoring}.

These trade-offs suggest that no frame of reference is universally optimal. \citet{belo2026userpreferencesuianchoring} found that usability of and preference for an anchoring strategy depends on the user's real-world activity and locomotion, the virtual and real-world tasks, the interaction modality and other properties of the UI element, and individual factors such as the user's mental model of where interfaces live. The same study reports that participants configured their anchoring once at the start of a condition and revised it in only 6 of 57 cases during the trials that followed, which suggests that the cost of reconfiguring was high relative to the benefit users expected from it. Anchoring is therefore a decision that a designer cannot make in advance and that a user is unlikely to revisit through manual controls. Whether speech can serve as a more seamless mechanism for making and revising this decision during an ongoing interaction has not been examined.

\subsection{Adaptive and Adaptable Interfaces}
Who decides how an interface changes is a long-standing question in HCI, and the two established answers are the system and the user.
\emph{Adaptive} interfaces automatically infer what to change and when. Studying adaptive menus, Gajos et al. found their perceived benefit and users' performance with such systems to largely depend on the accuracy of the  inference~\cite{gajos2006} and the predictability of the adaptation~\cite{gajos2008}.
\citet{findlater2004} compared static, adaptive, and adaptable menus and found that adaptable menus were faster than adaptive ones and preferred by most participants.
\citet{findlater2008} further showed that these effects depend on the interaction context, with adaptation becoming more valuable as the cost of accessing an element grows.

In \emph{adaptable} interfaces the user configures the behaviour directly, which removes the accuracy problem but introduces an interaction cost.
Early work by \citet{mackay1991} showed that customization is episodic rather than continuous: users change their software in response to specific triggers, and are held back by the time and attention the change demands.
\citet{banovic2012} demonstrated that this is partly a visibility problem, as surfacing customization opportunities in context substantially increases how often people take them up.
The same pattern holds in MR.
Direct manipulation techniques give users fine-grained, precise control over UI placement and anchoring~\cite{hayatpur2019spatialmanipulation, belo2026userpreferencesuianchoring, uimobility_pei2024}, but \citet{lu2022transition} found that users prefer semi-automated transition mechanisms over either fully manual or fully automatic ones, and the low reconfiguration rates reported by~\citet{belo2026userpreferencesuianchoring} indicate that the cost of manual customization is perceived as particularly high when users are engaged in time-critical tasks.

Our work explores a third approach.
Instead of the user configuring a static setting or the system continuously reassessing and inferring the user's needs from context, \system{} allows the user to iteratively refine through natural language how the interface should behave, and the system translates the user's intent into an adaptation policy which it refines or replaces as the user's needs change.
The user decides what the interface should do, while the system carries out that specification continuously. Whether users prefer this to configuring the interface themselves, and whether it lowers the effort of adjusting an interface during an ongoing task, is what we examine in this paper.

\subsection{Conversational Interaction with User Interfaces}
Natural language has been used to operate graphical interfaces for some time, most often to issue a single action in place of a manual one.
\citet{pasupat-etal-2018-mapping} ground commands such as \emph{``click on the second article''} to elements on a web page, and \citet{li-etal-2020-mapping} map multi-step instructions to action sequences on mobile UIs.
Large language models have widened what such systems can do: \citet{wang2023enabling} enable conversational interaction with mobile UIs, including question answering over a screen and mapping instructions to actions.
Language has also been used to interrogate interfaces rather than operate them, as by \citet{todi2021}, who let designers and developers query large GUI datasets conversationally.

Closer to our setting, recent work uses language to produce the interface itself.
\citet{malleableInterfaces} let end-users generate and iteratively tailor GUIs through natural language over an evolving task-driven data model.
In XR, \citet{multiagentic_li2026} allow users to open applications and specify their spatial layout through natural language commands, with clarification questions where the request is ambiguous.
In all of these cases the utterance resolves into a single outcome: an action is performed, an answer is returned, or a layout is produced.
The specification is complete once it is satisfied.

Our focus is different.
An anchoring decision is not a state to be reached but a rule about how the interface should behave from now on, and its appropriateness changes as the user moves, switches task, or is interrupted.
Speech is a promising interaction method for continuously refining the adaptive behavior of UI elements in MR because the hands and gaze are typically committed to the task that motivated the change in the first place.
We therefore study conversational interaction as a means of specifying and revising the ongoing anchoring and placement behaviour of MR UI elements, and how such interaction performs against and alongside direct manipulation during an ongoing task.

\section{Formative Study}
\label{sec:preliminary-study}

To better understand how users naturally describe spatial anchoring behavior in Mixed Reality and what a conversational system for controlling adaptive MR interfaces would need to support, we first conducted a formative qualitative study. Participants (N=19) interacted with an initial version of the study interface described in detail in Section~\ref{sec:user-study} 
(see Figure~\ref{fig:tasks}). Participant's primary task was to search for a specific key hidden in boxes placed throughout the room (simulated in an MR interface for better study control). The configuration of boxes changed throughout trials to enforce different locomotion conditions (stationary, semi-stationary, and mobile). While doing the task, participants interacted with three UI elements that differed in interaction modality, interaction frequency, and task-relevance and were encouraged to manually configure their preferred anchoring modes for each UI element using hand input and dragging the element (see more details below~\ref{sec:study-design}). 
After completing the tasks, we asked participants  the following open-ended question and recorded their responses:

\begin{quote}
\emph{``Instead of moving and anchoring the interface, imagine you could talk to it. How would you tell the interface to behave?''}
\end{quote}

\begin{table*}[t]
\caption{Themes identified from participant responses. Multiple themes could be assigned to a single response.}
\label{tab:preliminary_themes}
\centering
\small

\begin{tabularx}{\textwidth}{c l X c}
\toprule
\textbf{\#} & \textbf{Theme} & \textbf{Description} & \textbf{Count} \\
\midrule

1 & Reference Points &
Specifying locations relative to the body, environment, or other interfaces.
\newline\textit{Example: ``Key task appear next to the workstation.''}
& 55 \\

2 & Layout Management &
Describing spatial arrangements among multiple interfaces.
\newline\textit{Example: ``Align all the UIs horizontally.''}
& 17 \\

3 & Anchoring Related &
Defining how interfaces should be anchored, such as world, or body anchoring.
\newline\textit{Example: ``Attach to my left arm.''}
& 21 \\

4 & Visibility Rules &
Controlling when interfaces should appear, disappear, or hided.
\newline\textit{Example: ``Appear/Hide when I want.''}
& 18 \\

5 & Rule-Based Triggering &
Triggering interface behavior based on events, requests, or task conditions.
\newline\textit{Example: ``Show only when I receive a call.''}
& 12 \\

6 & World / Ergonomics Knowledge &
Requests requiring the system to account for ergonomics, surrounding objects, or task context.
\newline\textit{Example: ``Show in front of me without obstructing my view of the tasks I need.''}
& 13 \\

7 & Movement / Gaze Based &
Requesting interfaces to follow user movement, pose, gaze, or attention.
\newline\textit{Example: ``Follow me.''}
& 27 \\

8 & Adjustments &
Fine-grained modifications such as repositioning, or correcting placement.
\newline\textit{Example: ``Go a bit further.''}
& 17 \\

9 & Interface Summoning &
Explicitly referring to individual interfaces or requesting a way to access them.
\newline\textit{Example: ``I'd like to find some way to call each interface.''}
& 42 \\

10 & Deviation Constraint &
Maintaining previous configurations and limiting undesired changes.
\newline\textit{Example: ``Do not move too much from the place I put it.''}
& 5 \\

\bottomrule
\end{tabularx}

\end{table*}

We collected 57 responses across the three locomotion conditions. Each response described how participants would request adaptations for each of the three UI elements using natural language, specifying the different interaction needs and preferences arising from the corresponding locomotion condition and the different interaction requirements of each UI element. The format of responses was mixed with some formulating actual commands (e.g., "Attach to my left arm") while others were more descriptive (e.g. "I'd like to find some way to call each interface"). The responses were coded inductively in a bottom-up process using ten thematic labels by three researchers from this project with the goal to capture the requirements for such an interface. Since a single utterance could express multiple adaption intents, multiple labels could be assigned to the same response. In total, the coding produced 245 label assignments across the 57 responses.

Table~\ref{tab:preliminary_themes} summarizes the identified themes, their frequencies, and example utterances. The most prevalent theme was \textit{Reference Points} (\#1, $n=55$), indicating that most of the participants described the interface placement relative to locations on the body, in the environment, or with respect to other interface elements. \textit{Interface Summoning} (\#9, $n=42$) was the second most frequent theme, showing that users often explicitly called the target interface and requested to summon it in front of them. \textit{Movement / Gaze Based} (\#7, $n=27$) and \textit{Anchoring Related} (\#3, $n=21$) were also common, which could highlight the importance of both dynamic behaviors (like "following the user") and persistent spatial relationships (like attaching interfaces to the body or environment). Visibility management (\#4, $n=18$), layout organization (\#2, $n=17$), and adjustment commands (\#8, $n=17$) show that users intuitively specify both high-level spatial behaviors and positions, as well as fine-grained refinements.
The remaining themes reveal additional expectations, like ergonomic considerations (\#6, $n=13$), such as reachability and avoiding visual obstruction, and occasional rule-based specifications formulated by participants (\#5, $n=12$), such as displaying a specific UI element only when receiving a call. A smaller number of responses expressed constraints on preserving previous placements or limiting larger deviations (\#10, $n=5$). 

Overall, our themes show that natural language commands for MR adaptation are compositional and can combine several criteria and targets in a single utterance. We note that participants answered in an interview, with time to think and without seeing the resulting interface behavior, so the responses are one-shot descriptions rather than the turns of an ongoing interaction. In practice, users might not specify all desired adaptations within a single utterance but take multiple turns to refine their adaptation intent. Moreover, a user's message might be ambiguous and underspecified, for example, not specifying parameters (e.g. how far away something should be placed), leaving trade-offs unresolved (e.g. is the visibility of the UI or the real-world task more important if both cannot be fulfilled) or not mentioning relevant criteria (e.g. the UI element should be visible \emph{and} reachable). A user might forget to specify them or not be aware of it before experiencing the resulting UI behavior. Importantly, adaptation needs will change as the user's task or environment change, requiring users to periodically revise or adjust previously specified behaviors. Finally, since intent extraction is inherently uncertain, the system may misinterpret a command or realize it in a way the user did not anticipate. Users therefore need to understand what was applied and be able to revert it.

From these findings and our reflections, we define five design requirements for a conversational system for MR adaptation that guide our design of \system{} which we develop in Section~\ref{sec:system}:

\begin{itemize}

\item[\textbf{DR1}] \textbf{Compositional commands.} Interpret messages that jointly specify a target interface, spatial references, desired behaviors and constraints within a single utterance.

\item[\textbf{DR2}] \textbf{Grounded reference resolution.} Allow explicit and implicit references to interface elements, body parts, physical surroundings, virtual objects, or previous adaptations.

\item[\textbf{DR3}] \textbf{Persistent and conditional behavior.} Treat a command as a standing policy. Continuously re-evaluate it as the user's pose, task, and environment change, and trigger adaptations on specified conditions.

\item[\textbf{DR4}] \textbf{Handling of underspecified commands.} Infer suitable parameters from environmental and ergonomic context when a user states a goal without the means to achieve it. Ask for clarification only when the interpretation is uncertain and might lead to undesired behavior.

\item[\textbf{DR5}] \textbf{Refinement, revision, and transparency.} Support incremental refinement, replacement, and reversal of previous adaptations across turns, and communicate which adaptations were applied.
\end{itemize}

\section{System Design}
\label{sec:system}
\begin{figure*}[t]
\centering
\includegraphics[width=1\linewidth]{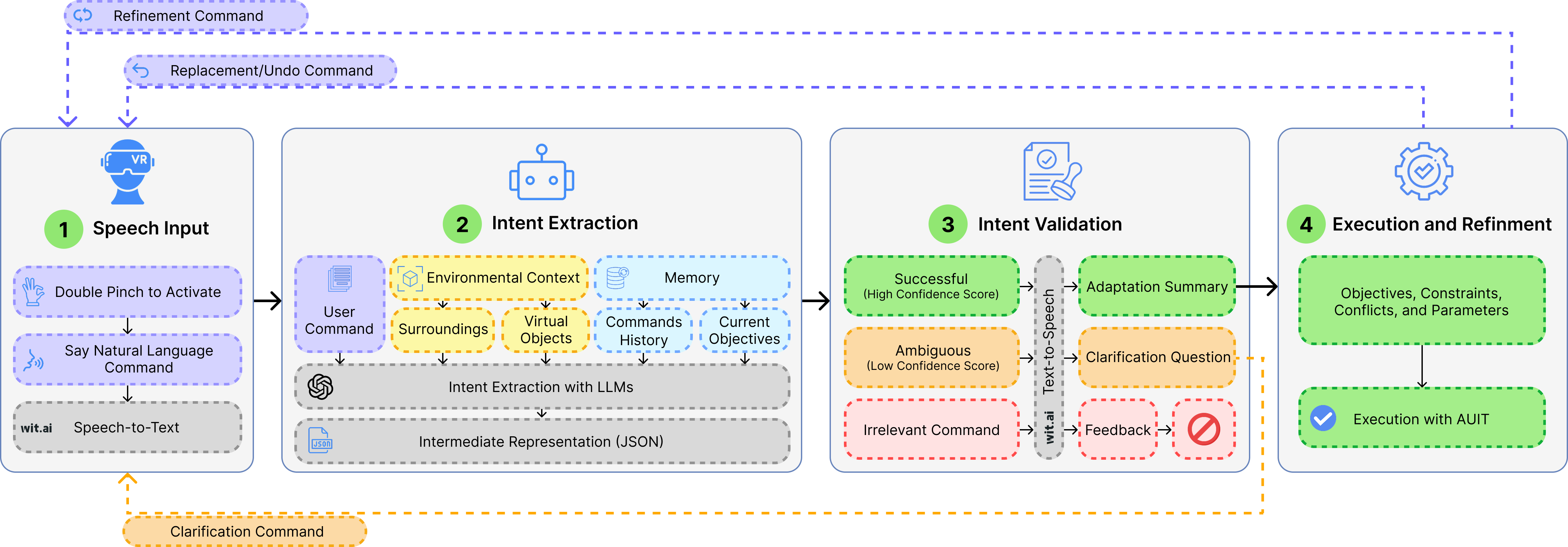}
\caption{Overview of the \system{} pipeline, comprising four stages. A spoken command (1) is interpreted against the environmental context and the interaction history into an intermediate representation (2), which is either
forwarded for execution, clarified, or rejected (3). Execution yields a set of weighted objectives, constraints, and parameters that the interface satisfies continuously while the user performs their task (4) and which users can refine or replace.}
\label{fig:system-overview}
\end{figure*}

We next develop \system{}, a conversational interface for controlling MR interface adaptations through natural language. \system{} allows a user to speak a request, interprets the user's utterance against the current adaptation policies and environment, and translates it into a new adaptation policy. 
In the following, we describe the four steps of this pipeline, visualized in Figure~\ref{fig:system-overview}, and how they address our design requirements. 

\subsection{Speech Input}
\label{sec:speech-input}
The user activates \system{} through a double pinch and then speaks an instruction describing the desired adaptation. We use an explicit activation gesture rather than continuous listening to avoid unintended activation, since our study tasks also required participants to speak. The utterance is transcribed with a speech-to-text module (Meta Voice SDK). We treat the resulting transcript as noisy rather than attempting to improve recognition itself and handle recovery in the next steps. .
 
\subsection{Intent Extraction}
\label{sec:intent-extraction}
The goal of this step is to turn the transcript of the user's utterance into a specification for UI adaptation that can be executed. Commands collected in the  formative study show that such utterances can be  expected to be  compositional. They will jointly name a target, a spatial reference, a desired behavior, or constraints (DR1). They might also refer to entities in the world and to earlier commands (DR2).

We therefore pass the transcript to an LLM (OpenAI Responses API, GPT 5.2) and require a JSON output following a fixed schema we devised. Along with the transcript, we also pass a structured scnapshot  of the current scene, containing the available interface elements, relevant objects and surfaces in the environment (through MRUK), and their spatial information such as positions and bounds, as well as the user's current position as a structured textual representation of the current context. We are also passing the recent commands, conversational turns, and current adaptation state, which allows the model to resolve pronouns and contextual expressions such as ``move it higher'' or ``make that follow me'', and revert or adjust the previous adaptations.
 
The JSON representation specifies the target interface, the adaptation objectives, spatial references and relations, body-relative placement, an interpretation confidence, clarification question (if applicable - for ambiguous commands), and whether the system should apply, clarify, or undo an adaptation. Specifically, adaptations are expressed through predefined objectives, based on those provided by  AUIT~\cite{belo2022auit}, including visibility, anchoring, positioning, keeping interface elements in view, and spatial movement. In addition, we support distance constraints, reference points, and relative movement commands. It can also represent multiple objectives for multiple interface elements. 

The prompt template was developed from the command set identified in Section~\ref{sec:preliminary-study} and refined through iterative testing with variations of those commands, checking whether the model could robustly interpret paraphrases, incomplete utterances, follow-up instructions, and noisy speech-recognition outputs. It is described in detail in Appendix \ref{app:ir}. Two parts of it are specific to controlling adaptive behavior. We introduced conflict-resolution rules to determine whether a new adaptation is merged with the interface's existing behavior or replaces a conflicting prior constraint, so that conflicts are settled or flagged during interpretation for the next steps. Rules for conditional behavior let a command apply only while the user is moving, turning their head, looking at an interface, or while a task is active (DR3). The prompt further instructs the model to treat transcripts as noisy and to correct likely misheard words against the scene entities and interface names, and includes few-shot examples for these difficult cases. Importantly, it asks the model to assess its own confidence in the interpretation, which informs the next step.

\subsection{Intent Validation}
\label{sec:intent-validation}
 
Both the transcription of the users utterance and the interpretation of it can fail or contain mistakes, and an adaptation the user did not ask for is costly in MR because it can obstruct the real-world task. Thus, rather than requiring a confirmation or correction for each command, we make the system's interpretation visible and ask  clarification questions only if the system's confidence is low  (DR4, DR5).
 
To this end, the intent validation steps inspects the confidence value reported in the intermediate representation to determine whether and how to proceed with the interaction. If confidence is high (above 80\%), the adaptation is applied and \system{} reports a summary of the detected adaptation through text-to-speech (Meta Voice SDK / Wit.ai) and as written text, so that users can relate the observed interface change to what the system understood. This summary is purely informational. Users can undo the adaptation at any point. If confidence is low or multiple interpretations are possible, the system instead asks a follow-up question, which is indicated in the clarification options in the intermediate representation, before applying anything, and the user's answer re-enters the pipeline as a new turn. If the command falls outside the supported adaptation domain or is unrelated to interface adaptation, the system outputs feedback to that effect and makes no change.
 
\subsection{Execution and Refinement}
\label{sec:execution}
The adapation needs verbalized by users have to be turned into standing adaptation policies that have to hold as the user's pose, task, and environment change (DR3), and that the user must be able to revise once they have seen its effect (DR5).
 
The final step therefore maps the structured representation to adaptation constraints aligned with objective categories defined by AUIT and translates it into executable adaptation rules applied to the respective UI element, to update its spatial position in real time while the user continues their task. \system{} then performs multi-objective optimization using AUIT's weighted objective costs. 
We update objective weights from the recent conversational history through a recency-based deterioration factor, so that recently requested adaptations receive higher priority at runtime and users can override earlier requests as their needs change. Concretely, for each incoming command, the weights of previously stored commands are multiplied by a deterioration factor of 0.95, gradually reducing the influence of older commands over time. We keep two objectives always active, independently of what the user requests to avoid occlusion of the user's view and to keep interfaces oriented toward the user (same for the direct manipulation baseline, see Section~\ref{sec:dm}).
 
Upon seeing and hearing about the effect of their command, users can engage in a refinement or replacement loop, which uses the same mechanism. They can modify, extend, undo, or clarify previous adaptations in additional turns, grounded in the memory module described in Section~\ref{sec:intent-extraction}, so that the interface can be adjusted iteratively and does not need to be configured in one shot or reconfigured from scratch.

\section{User Study}
\label{sec:user-study}

We conducted a mixed-methods user study to investigate how conversational interaction can support interface adaptation in mixed reality. In a within-subject study, participants went through three blocks of trials where they used, \system{}, a direct manipulation (DM) baseline and a combination of both (Combined) to customize the adaptive behavior of three UI elements with different interaction properties and in response to changes in their  task and environment. 
In particular, we were interested in exploring the following research questions: 

\begin{description}
  \item[RQ1] How does conversational control of MR interface adaptation  compare to direct manipulation in terms of  usability and preference?
  \item[RQ2] Which factors lead users to prefer one interaction method over the other, and what trade-offs do they perceive between them?
  \item[RQ3] How does the availability of conversational control change what and how often users customize MR adaptations?
  \item[RQ4] How do the adaptation needs expressed in live interaction compare to those elicited in the formative study, and how well does \system{} support them?
\end{description}

In the following, we first describe the DM baseline and experimental design, followed by the study task, procedure, participants, and apparatus. We then describe the data we collected and our analysis approach.

\subsection{Direct Manipulation Baseline}
\label{sec:dm}

\begin{figure*}[t]
    \centering
    \includegraphics[width=\textwidth]{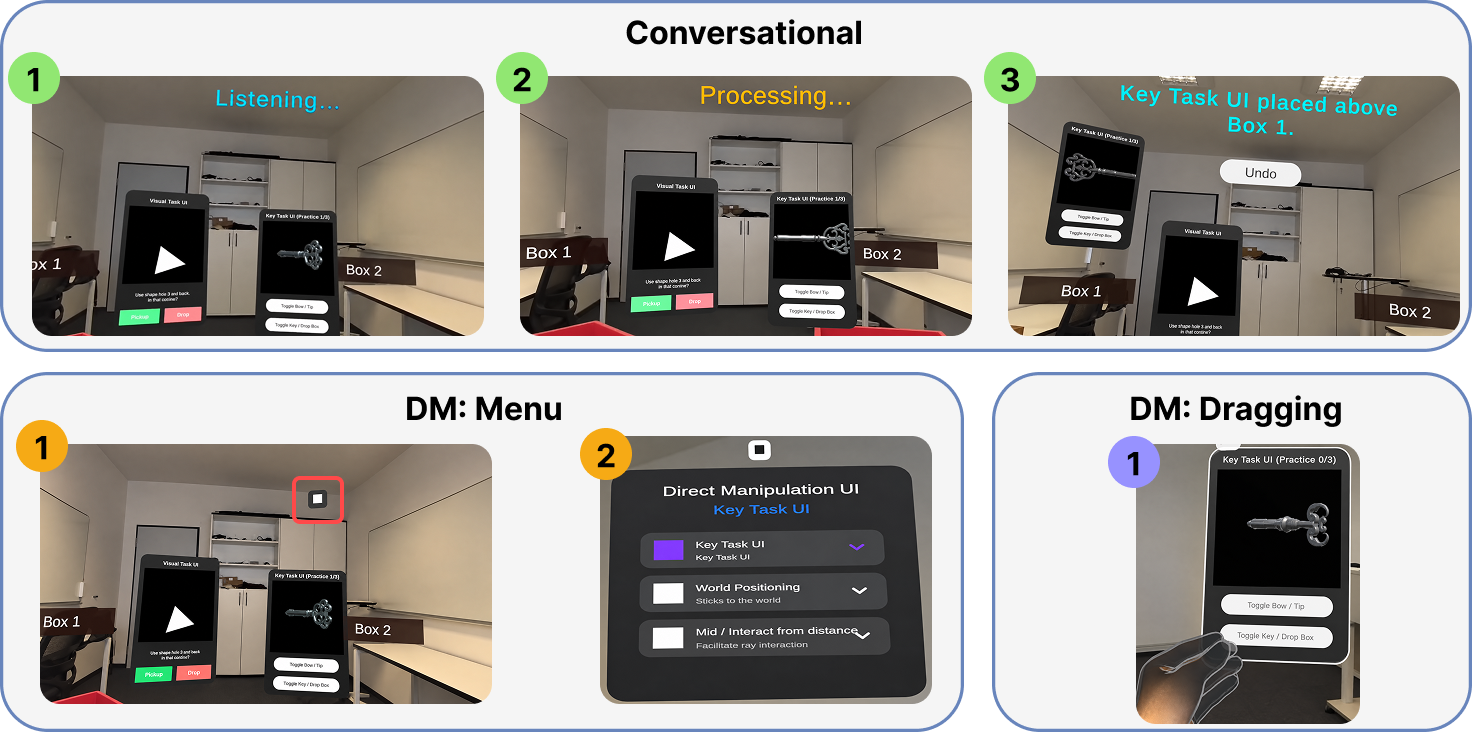}
    \caption{
    Interaction methods used in the study.
    \textbf{Conversational (top):}
    \fignum{convgreen}{1} after activating \system{}, the system entered
    listening mode and participants issued a spoken adaptation request;
    \fignum{convgreen}{2} \system{} processed the request, and
    \fignum{convgreen}{3} the resulting interface adaptation was applied. \system{} provided written and spoken feedback, and users could undo the changes.
    \textbf{DM: Menu (bottom left):}
    \fignum{dmorange}{1} participants opened the head-anchored DM menu using
    the access button (it is highlighted using a red box around it), and
    \fignum{dmorange}{2} selected an interface and configured its adaptation
    properties through the menu.
    \textbf{DM: Dragging (bottom right):}
    \fignum{dragpurple}{1} participants directly grabbed and repositioned an
    interface using hand-based dragging.
    }
    \label{fig:interaction-methods}
\end{figure*}

We implemented a Direct Manipulation (DM) baseline, which combines spatial placement through hand-based dragging with a graphical control panel for selecting a predefined set of  adaptive behaviors, that allowed users to perform almost the same types of interface adaptations supported by \system{} through hand input manually or using a menu with a set of configurations for each interface. 
Figure~\ref{fig:interaction-methods} illustrates the interaction flow for conversational control and the
two forms of direct manipulation available in the DM baseline.
Users can directly grab and reposition interface elements to make continuous spatial adjustments. For configuring adaptation behaviors, DM provides a compact, head-anchored menu containing separate controls for the three adaptable interfaces in the environment. Users first click on a small access button to open the menu, select an interface they want to adapt and can then modify its adaptation properties by choosing a frame of reference (head, torso, hand, or world-anchored) and an interaction distance (near, mid, or far). All UI elements automatically avoid occlusions and maintain orientation toward the user.
Changes selected through the panel are applied immediately and can subsequently be refined either through further interface selections or by directly grabbing and repositioning the interface.

The DM menu was intentionally kept simple and exposed only anchoring and distance controls, more than most prior work (e.g.,~\cite{belo2026userpreferencesuianchoring}). Earlier iterations also included options such as grid-based positioning relative to anchors, visibility settings, and occlusion management. However, pilot testing showed that adding these controls quickly increased menu complexity and interaction time.
We thus made them accessible through manual dragging.

\subsection{Study Design}
\label{sec:study-design}

We used a within-subjects design with one independent variable, \textsc{Interaction Method}, comprising three conditions: DM, \system{}, and Combined. In the DM condition, participants adapted the interfaces using the graphical panel and direct spatial manipulation described above. In the \system{} condition, participants performed adaptations through spoken natural-language commands as descibred in~\autoref{sec:system}. In the Combined condition, both interaction methods were available.

Participants conducted 3 blocks of 6 trials each, one for each condition. The order of DM and \system{} was counterbalanced across participants, with 12 participants completing DM first and 12 completing \system{} first. Combined was always presented as the third block, ensuring that participants had prior experience with both individual interaction methods before deciding how to use them together.

\subsection{Experimental Tasks}
\label{sec:tasks}
Our experiment leverages a dual-task paradigm with one primary task and two secondary tasks with three UI elements overall, closely following previous work~\cite{belo2026userpreferencesuianchoring}. This reflects a common use of MR interfaces which allow people to seamlessly switch between real-world tasks and digital content \cite{track_rasch2025}. The three UI elements corresponding to each task require different interaction modalities and frequencies and the set up of the primary task demands different levels of movement throughout the world. Together, these mimic frequent changes in a user's task and environment and participants were asked to adapt the spatial placement and adaptation behavior of UI elements to prioritize their comfortability over speed and accuracy, allowing them to interact with the UIs while busy with the tasks comfortably. To make spatial references easier during conversational interaction, we added visible labels above all boxes and other virtual objects, and displayed the name above each interface, allowing participants to refer to them directly in their commands (e.g., ``put the Key Task UI above Drop Box 1'').

\begin{figure*}[t]
    \centering
    \includegraphics[width=\textwidth]{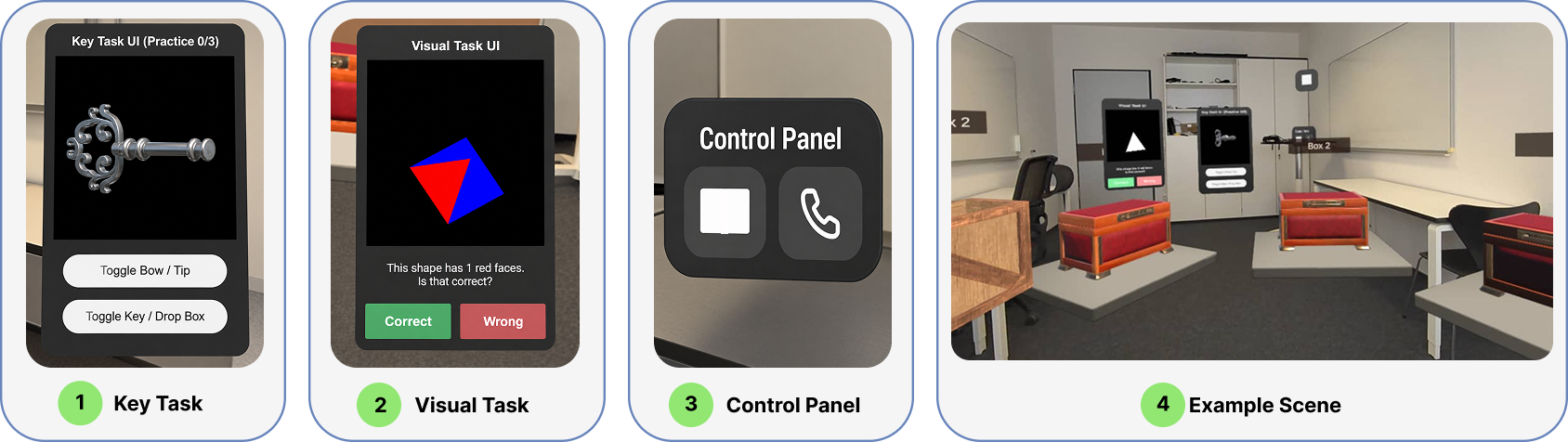}
    \caption{The three adaptable interfaces used in the study:
    (1) the Key Task UI, which provided instructions for the primary
    task and required visual and manual interaction;
    (2) the Visual Task UI, used for the incoming visual task; and (3) the Control Panel, used for brief manually controlled
    system events such as calls or music playback;
    (4) a sample scene of the boxes and the interfaces.}
    \label{fig:tasks}
\end{figure*}

The primary task in each trial was the \emph{Key task}: participants searched for the correct key among five alternatives in boxes placed throughout the room, took it out of the box and put it into one of two delivery boxes. An instruction interface (see \autoref{fig:tasks}) displayed the correct handle and tip of the key, as well as the intended delivery box, which participants needed to toggle through using buttons. The Key task UI thus required both visual attention and manual interaction. The spatial arrangement of the  boxes was used to induce different mobility levels.
In the first set, the boxes were positioned around the participant and could be reached without substantial movement. In the second one, boxes were grouped across multiple locations in the room, separated by virtual walls, requiring participants to move between areas. In the third one, all boxes were distributed around the room, requiring continuous movement through the environment. Each arrangement was repeated for two trials with the order randomized within each \textsc{Interaction Method} block.

At random intervals (5-15 seconds after the start of the key task, and again 5-15 seconds after completing the first secondary task),, participants were prompted to do one of two secondary tasks. 
In the \emph{Visual Task} (see \autoref{fig:tasks}), participants monitored a continuously rotating shape with differently colored faces. At random intervals, they were prompted to report verbally how many faces matched a specified target color. The task required sustained visual attention only during short intervals but no manual interaction, representing a glanceable background-monitoring interface. In the  \emph{Control Task}, participants responded to intermittent simulated system events, such as an incoming call or music playback, using hand-poke controls (see \autoref{fig:tasks}). This task represented brief, manually controlled interruptions that required the corresponding interface to remain accessible during other activities.

\subsection{Procedure}
\label{sec:procedure}

Each session began with informed consent and a background questionnaire. The experimenter then introduced the purpose of the study and the concept of MR interface adaptation using a cooking scenario as a motivating example. Participants were introduced to the three tasks, the DM panel and grabbing interaction, and how to address \system{}. To familiarize participants with the capabilities of the potential adaptations without prescribing particular commands, the experimenter provided a high-level description of the supported adaptation categories without giving example utterances.

Participants first completed a Key-Task-only familiarization round to become comfortable with basic MR interaction. They then completed three practice trials, one with each mobility level. Both DM and \system{} were available during practice, and participants were encouraged to try direct grabbing, the graphical panel, and conversational interaction. 
After familiarization, they conducted 3 blocks of 6 trials each, as described above.
Before each trial started, participants were given the chance to familiarize with the current placement of boxes and to configure their intended adaptation over the three interfaces. When ready, they could press a button to start the trial. This mechanism ensured that they were not in a rush by the task, and they had enough time to prioritize their interaction comfort over task execution time and to apply adaptations based on the mobility level, task characteristics and personal preference.
After each block, participants completed a questionnaire followed by a semi-structured interview about their experience with that \textsc{Interaction Method}. After completing all three blocks, participants completed a final questionnaire and closing interview, followed by a debriefing.
The study lasted 90 minutes per participant and was rewarded with 20€. An experimenter was present throughout the session.
The study protocol was reviewed and approved by the university's ethics committee.

\subsection{Participants}
\label{sec:participants}

We recruited 24 participants (13 male, 11 female)
through advertisements on the university campus and word of mouth. Participants were between 21 and 31 years old ($M=24.6$, $SD=2.3$) and reported relatively limited prior experience with VR/AR/MR ($Mdn=2$, $M=2.5$) and moderate videogame use ($Mdn=3$, $M=3.5$). Familiarity with LLM-based tools was high ($Mdn=7$, $M=6.7$), whereas use of voice assistants was more moderate and varied ($Mdn=3$, $M=3.5$). All four measures were assessed on 7-point scales.

\subsection{Apparatus}
\label{sec:apparatus}
The study was conducted using a Meta Quest 3 headset in an approximately $3.5 \times 5$,m indoor study space. Passthrough was enabled throughout the study, allowing participants to perceive and move through the physical environment while interacting with the virtual interfaces. Participants interacted with the interfaces using the headset's optical hand tracking; no handheld controllers were used.

The study application was developed in Unity (version 6000.3.6f1) and implemented the DM method described in \autoref{sec:dm}, the three tasks described in\autoref{sec:tasks}, and the \system{} conversational adaptation pipeline described in Section~\ref{sec:system}. For enabling the adaptive behavior, we incorporated the AUIT toolkit ~\cite{belo2022auit}.
We used Claude as an AI-assisted programming tool during development to support code generation, debugging, and code refinement; all generated code was reviewed, tested, and integrated by the authors.

\subsection{Data Collection and Analysis}

\subsubsection{Data Collection}
\label{sec:data-collection}
We collected quantitative, qualitative, and behavioral data throughout the study. Quantitative data on user experience, ergonomics, and usability of each \textsc{Interaction Method} was collected through structured questionnaires after each trial block 
containing eight Likert-scale items. Preferences towards each method were collected through two bipolar comparison items in the post-study questionnaire  (see items in Table~\ref{tab:likert} and Appendix~\ref{app:questionnaire}). 
Qualitative explanations and preferences were collected through two open-ended questions after each \textsc{Interaction Method}, and questions about perceived trade-offs and situations in which participants would prefer each interaction method at the end of the study. Questionnaire items were adapted from validated instruments where a close match existed and written as study-specific items otherwise; the complete item wording and provenance are provided in ~\ref{app:questionnaire}.
Open-ended questions were administered as semi-structured interviews rather than as typed responses. The experimenter asked each question and followed up conversationally where appropriate. During the sessions, the experimenter also recorded voice from the sessions to write transcripts, and also kept observational notes about participants' interactions, comments, difficulties, and other events relevant to their experience. These notes were used alongside participants' interview responses in the subsequent qualitative analysis.

The application additionally logged participants' interactions throughout each session. The logs captured session and trial events; the conversational pipeline, including speech transcription, the LLM-generated intermediate representation, and the applied adaptation; DM and adaptation-history events, including undo states; and performance on the three secondary tasks. These logs enabled us to reconstruct participants' adaptation behavior and analyze the conversational pipeline from spoken input to executed interface change.

\subsubsection{Data Analysis}
\label{sec:data-analysis}

\paragraph{Questionnaire Data.}
To answer RQ1, We analyzed quantitative data from questionnaires to examine participants' perceived usability and experience across the interaction methods. We compared Likert-scale responses across interaction methods using Friedman's test, followed by Holm-corrected Wilcoxon signed-rank tests for pairwise comparisons where appropriate. Because Likert responses are ordinal and our inferential analyses are rank-based, we use the median as the primary measure of central tendency and report means alongside it to facilitate comparison with prior work. Categorical responses are reported as counts and percentages. 

\paragraph{Qualitative Data.}
To answer RQ2, We conducted an inductive thematic analysis of the qualitative data. One researcher, who also conducted the study sessions, performed the coding. The analysis drew on participants' responses from the Post-Method and Final interviews together with the researcher's observational notes and notes from discussions during the sessions. The researcher reviewed each participant's qualitative data holistically and developed codes inductively from recurring patterns in the data rather than using a predefined coding scheme. Codes were assigned based on the content of participants' responses rather than the specific question that elicited them, as participants frequently raised issues relevant to multiple or adjacent questions. The codes were iteratively reviewed and grouped into higher-level themes capturing recurring experiences, preferences, and trade-offs across the interaction methods.
We used an LLM (Claude) as an assistant during this process to help organize codes and suggest potential groupings and themes; all resulting themes were reviewed, refined, and finalized by the researchers.

\paragraph{Log data.}
We analyzed the interaction logs to answer RQ3 and RQ4. For RQ3, we computed customization frequency per participant and condition, counting each issued voice command or DM menu selection as one adaptation action. Hand-based grabbing and dragging were not logged as discrete actions and are therefore excluded from this measure, making DM and Combined frequencies a lower bound on manual activity. For RQ4, we analyzed conversational commands at the individual-command level to characterize participants' adaptation needs, compare them with the themes from the formative study, examine how commands were clarified, refined, replaced, repeated, or reverted over time, and evaluate the different stages of the conversational pipeline.

\section{Results}

We report our results following the research questions introduced in Section ~\ref{sec:user-study}. To separate the different forms of evidence, we first present the quantitative results from participants' ratings and preferences, followed by the qualitative findings explaining the trade-offs behind these preferences. We then report the log-based analyses of natural-language commands and customization behavior. Unless noted otherwise, Likert items use 1-7 scales and are reported as $Mdn$ with $M$ alongside; statistical procedures are described in ~\ref{sec:data-analysis}.

\subsection{Quantitative Results}

\subsubsection{Subjective Usability and Preference}
\label{sec:results-quant}
\begin{figure}[t]
\centering
\includegraphics[width=\linewidth]{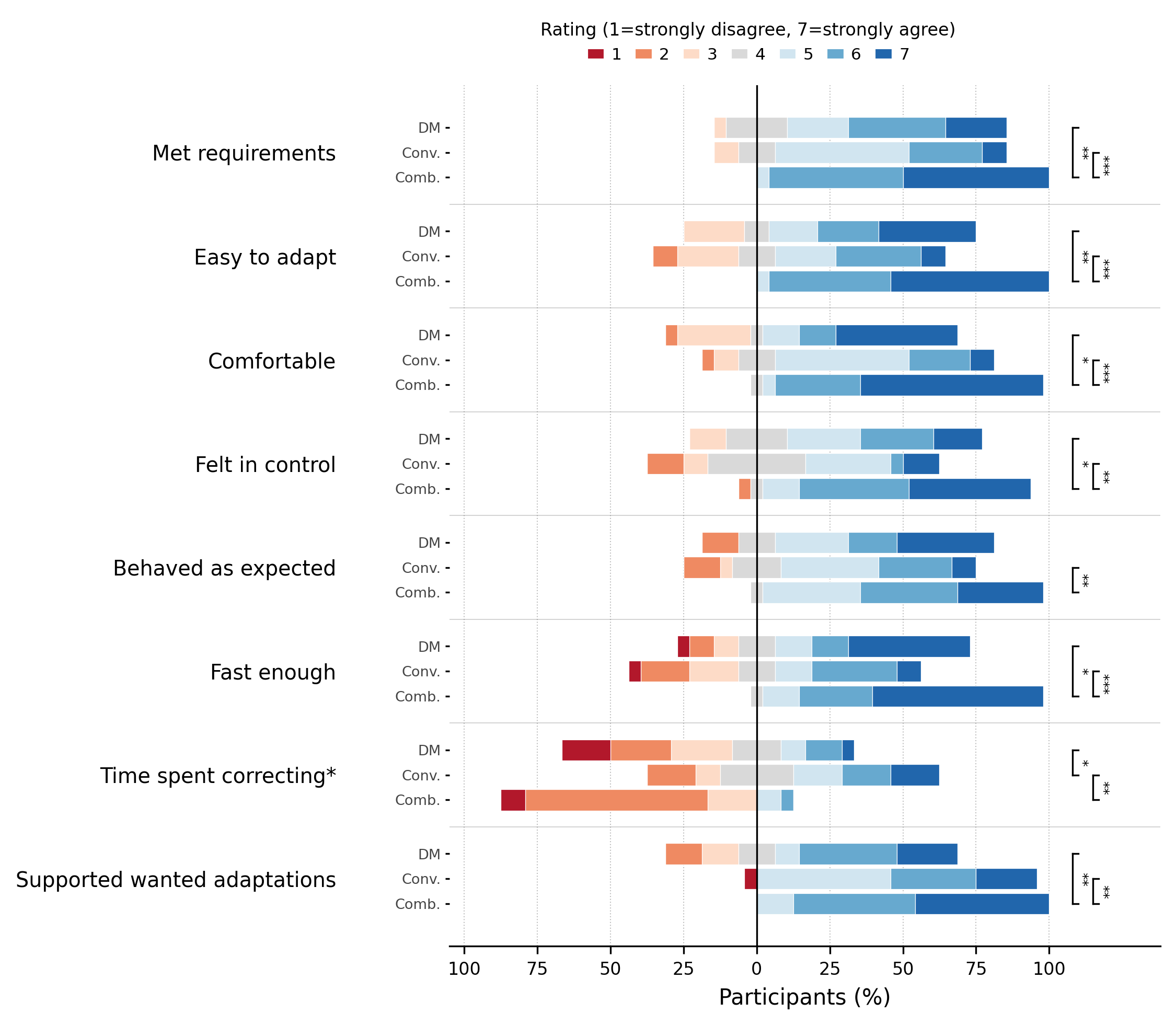}
\caption{Post-block Likert ratings (1–7, higher = better) by condition, shown as diverging stacked bars. Rows within each item are DM, ConverseMR (Conv.), and Combined (Comb.); Brackets on the right mark pairwise differences that remained significant after Holm correction (*p < .05, **p < .01, ***p < .001). *Reverse-worded item (higher = worse). Numeric values in Table 2.~\ref{tab:likert}.}
\label{fig:likert-boxplot}
\end{figure}

As visualized in Figure~\ref{fig:likert-boxplot}, Combined was rated best  across all usability items with Friedman tests showing significant differences for all eight items 
(all $p \leq .012$; five at $p \leq .001$; see Table~\ref{tab:likert})
Pairwise comparisons show a consistent pattern: Combined significantly outperformed \system{} on all eight items and DM on six of eight, while DM and \system{} were statistically indistinguishable on seven of eight after Holm-correction. However, participants reported spending significantly more time correcting mistakes with \system{} than DM ($Mdn=4.5$ vs.\ $3.0$; $M=4.58$ vs.\ $3.33$; $p=.031$). 

This held at the level of overall preference, too: end-of-session ratings of how much participants would like to use each method going forward ranked Combined highest ($Mdn=7.0$, $M=6.62$, $SD=0.65$), DM second ($Mdn=5.0$, $M=5.17$, $SD=1.20$), \system{} lowest ($Mdn=4.5$, $M=4.29$, $SD=1.04$; $\chi^2(2){=}36.46$, $p<.001$).

Participants' account of \emph{when} each method wins matched a direct bipolar item almost perfectly: 23 of 24 rated DM better for simple, minor changes, and 21 of 24 rated \system{} better for complex and bigger adaptations. Isolating DM's two components shows neither alone explains its overall score: menu-only ($Mdn=3.0$, $M=3.05$, $SD=1.36$) and dragging-only ($Mdn=4.0$, $M=3.71$, $SD=1.55$) both scored well below full DM ($Mdn=5.0$, $M=5.17$, $SD=1.20$). 
Users argued that using the dragging without the complex behaviors in the menu was not ideal, and use of menu without subtle and customized adjustments manually was also problematic. However, dragging was not available in \system{}.

\begin{table*}[t]
\caption{Post-block Likert ratings (1--7) by condition, reported as $Mdn$ ($M$, $SD$) per Section~3.7. Friedman $\chi^2(2)$ across all three conditions. Pairwise columns show which conditions differed significantly after Holm correction (D=DM, C=\system{}, X=Combined).}
\label{tab:likert}
\small

\begin{tabular*}{\textwidth}{@{\extracolsep{\fill}}lccccc@{}}
\toprule
\textbf{Item} 
& \textbf{DM} 
& \textbf{\system{} (C)} 
& \textbf{Combined (X)} 
& $\boldsymbol{\chi^2}$ 
& \textbf{Sig.\ pairs} \\
\midrule
Met requirements
& 6.0 (5.46, 1.18)
& 5.0 (5.12, 1.03)
& \textbf{6.5 (6.46, 0.59)}
& 21.00***
& D-X, C-X \\

Easy to adapt
& 6.0 (5.38, 1.56)
& 5.0 (4.67, 1.52)
& \textbf{7.0 (6.50, 0.59)}
& 17.26***
& D-X, C-X \\

Comfortable
& 6.0 (5.29, 1.81)
& 5.0 (4.96, 1.20)
& \textbf{7.0 (6.50, 0.78)}
& 16.73***
& D-X, C-X \\

Felt in control
& 5.0 (5.12, 1.30)
& 4.0 (4.42, 1.44)
& \textbf{6.0 (6.04, 1.20)}
& 13.24**
& D-X, C-X \\

Behaved as expected
& 5.5 (5.33, 1.66)
& 5.0 (4.79, 1.44)
& \textbf{6.0 (5.88, 0.90)}
& 8.90*
& C-X \\

Fast enough
& 6.0 (5.25, 1.94)
& 4.5 (4.33, 1.81)
& \textbf{7.0 (6.38, 0.88)}
& 14.48***
& D-X, C-X \\

Time spent correcting$^r$
& 3.0 (3.33, 1.79)
& 4.5 (4.58, 1.69)
& \textbf{2.0 (2.50, 1.22)}
& 16.76***
& D-C, C-X \\

Supported desired adaptations
& 6.0 (5.00, 1.74)
& 5.5 (5.54, 1.25)
& \textbf{6.0 (6.33, 0.70)}
& 13.73**
& D-X, C-X \\
\bottomrule
\end{tabular*}

\vspace{2pt}
\begin{minipage}{\textwidth}
\footnotesize
$^r$Reverse-worded (higher = worse).
*$p<.05$, **$p<.01$, ***$p<.001$.
DM-\system{} never differed significantly except on the reverse-worded item.
\end{minipage}
\end{table*}

\subsection{Qualitative Results}

\subsubsection{Participants' Reasoning Behind Their Preferences}

We categorized participants’ reasoning about their preferences as  as described in~\ref{sec:data-analysis}. We identified the following themes.

\paragraph{Theme 1: Precision vs.\ Reach.}
One of the most common tradeoffs that participants described was between precise placement and the ability to make changes without physically reaching an interface. They appreciated DM for exact, close-range placement. As P13 explained, \emph{I can feel in control with fine positioning - drag it manually and put it EXACTLY where I want.''} In contrast, participants liked \system{} when interfaces were farther away or otherwise inconvenient to reach or they could not find the interface: \emph{You don't have to look for things - you can call for them and summon them''} (P7). This distinction is consistent with the direct bipolar item, where 23/24 participants favored DM for simple placement and 21/24 favored \system{} for complex placement. Distance also appeared in the interviews, with 18/24 participants identifying it as a condition that would push them toward \system{}.

\paragraph{Theme 2: Reliability vs.\ Capability.}
Participants highlighted a trade-off between the predictability of DM and the broader capabilities of \system{}. DM was perceived as reliable because its actions were directly controlled by the user, and deterministically users can position the interfaces: \emph{It is deterministic; I know what to expect''} (P13). In contrast, participants appreciated that \system{} could support a wider range of requests, but this flexibility also introduced the possibility of incorrect interpretation or execution. P7 described this frustration as: \emph{The language model... gaslights me - it says it is doing [the action], but it did not do it.''} 
This difference partly follows from our decision to limit the menu to anchoring and distance (Section~\ref{sec:dm-baseline}). P13 noted, ``I was really limited in positioning it, for example anchoring to the left hand, I could do that if the dropdown had more options.'' P16 pointed out the cost of the alternative, explaining that a longer menu would have pushed them toward conversational control anyway because finding an option by clicking became too time consuming. 

\paragraph{Theme 3: Visibility of Available Actions.}
Participants also differed in how clearly they could determine what each method supported. With DM, the menu made the available options visible, whereas participants did not always know which requests \system{} could understand or execute. P24 explained, \emph{Conversational - you don't know what it supports; DM - it shows all options.''} P17 described the same distinction as \emph{diversity (conversational supports a vast range of commands) vs.\ visibility (DM shows whatever it supports, but conversational does not).''} Thus, while conversational interaction provided flexibility in how participants could express requests, the boundaries of its capabilities were less visible. This concern also connects to participants' requests for clearer feedback about what the conversational system had understood and what action it had performed.

\paragraph{Theme 4: Physical and Mental Effort.}
Participants distinguished between the types of effort required by the two methods. DM required physical interaction, particularly when users needed to reach, move, or configure interfaces manually. As P20 noted, \emph{DM needs more physical effort, specifically with dragging.''} Conversational interaction reduced this physical effort, leaving the hands free for the ongoing task, which several participants valued: \emph{``I could do multitasking, while doing other tasks, it feels like adaptation happens meanwhile''} (P14).  However,  formulating an appropriate command felt effortful for some. P13 explained, \emph{I have to think about the prompt... waste a couple of seconds... to figure out how I can say command to a stupid machine.''} 

\paragraph{Theme 5: Speed.}
Participants' perceptions of speed depended on the type and complexity of adaptation being performed. DM was perceived as faster for simple, subtle actions and fine adjustments, whereas \system{} was perceived faster for distant or multi-step actions. For example, P17 explained that \emph{``conversational is faster in sense of time, since for anchoring you need to spend time for few clicks and steps to do it with the DM menu vs.\ reliability, which DM is more reliable.''} 
The same applied when several interfaces had to change at once, which participants could request in one utterance but had to configure one by one through the menu.
At the same time, other participants favored DM in terms of speed, and P19 considered the two methods approximately equal. 

\paragraph{Theme 6: Privacy and Social Context.}
Participants' preferences were also shaped by the social context of interaction. Although conversational interaction was useful when an interface was difficult to reach, speaking commands aloud could be less suitable in public or privacy-sensitive settings. P12 noted that \emph{``DM can be used publicly - you don't have to say everything out loud.''} while P20 independently identified privacy as a drawback of \system{} in  two comments.

\paragraph{Summary:} These themes describe factors participants weighed rather than an ordered set of preferences. \emph{Distance, adaptation complexity, and multitasking} pushed them toward \system{}, with 18 of 24 naming distance as a reason to speak rather than reach, 21 of 24 rating \system{} better for complex adaptations, and 23 of 24 rating DM better for simple ones. \emph{Precision, determinism, and visibility} of available options pushed them toward DM, as did settings where speaking aloud felt inappropriate. 

\subsubsection{Benefits and Costs of Having Both Methods Available}
Participants who preferred Combined described its benefit in near-identical terms, using conversational control for rough or distant changes and DM for close and exact ones (roughly 20 of 24). Several also valued having a fallback when one method failed. As P13 put it, \emph{``I was more confident that I had several methods to influence the interface. I knew that when the agent doesn't understand me, I have a backup.''}
Having both methods available was not free, although only a few participants raised a cost, and since these remarks were unprompted the counts below are lower bounds. Four participants described the effort of deciding which method to use for a given adaptation. P15 called it \emph{``decision fatigue because I wasn't sure whether conversational or DM was better for each task''}, and P24 reported the same. Three participants experienced cross-modal interference, accidentally triggering conversational control while manipulating an interface by hand or the reverse (\emph{``by mistake I activated the agent [while working with the menu]''}, P13).


\subsection{Log-Based Analysis}
Based on the collected log data described in ~\ref{sec:data-collection}, we analyzed the frequency of customizations defined across the different \textsc{Interaction Methods}. We then analyzed the formulated voice commands in terms of the themes identified in the formative study to understand how participants communicated their adaptation needs through voice input. Finally, we analyzed how the conversational interaction unfolded over time and used the collected data to validate the different parts of our \system{} pipeline.
Regarding the latency of the conversational pipeline, it had an end-to-end response latency with a median of 4.45 seconds across all commands for all 24 participants.

\subsubsection{Customization frequency}
\begin{figure}[t]
\centering
\includegraphics[width=\linewidth]{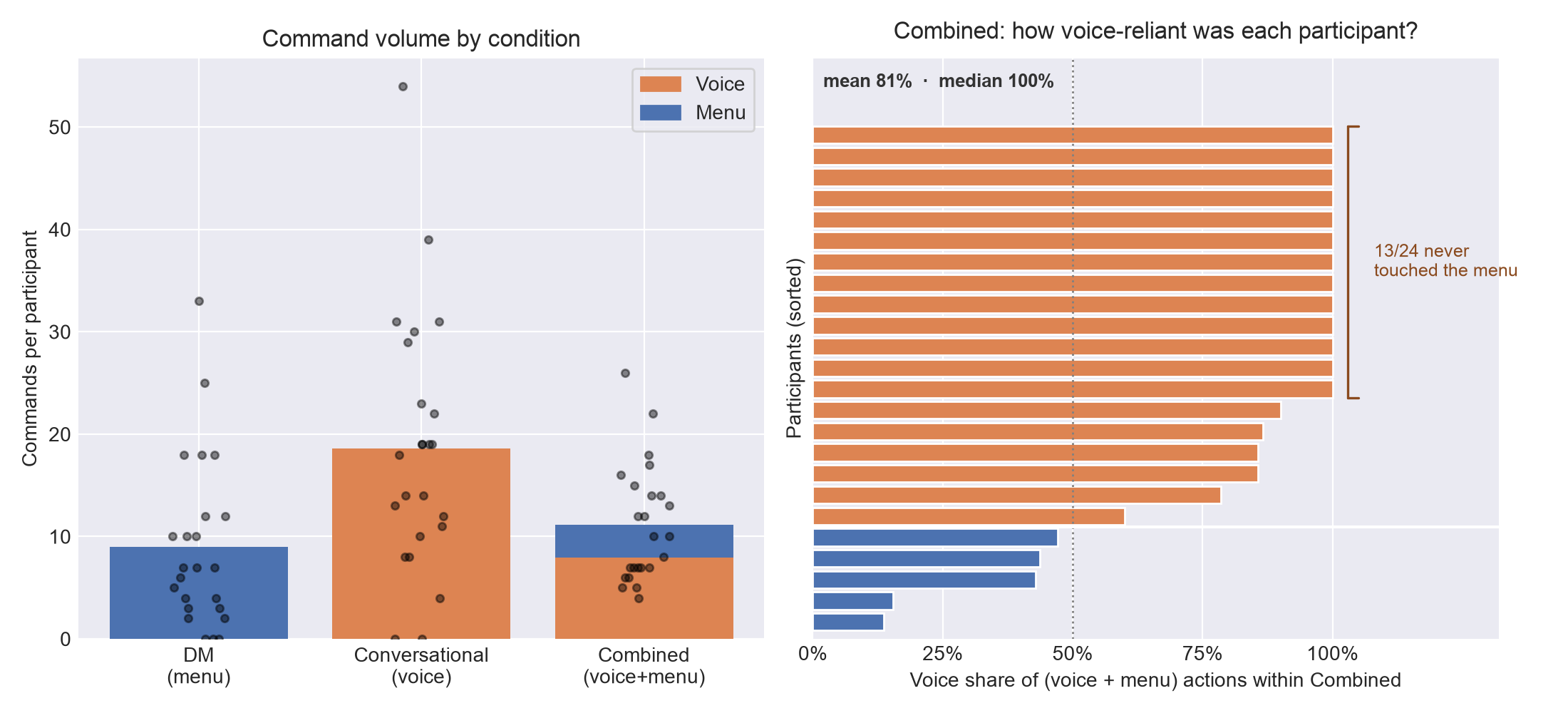}
\caption{Left: mean commands/interactions issued per participant by condition (Combined shown stacked, voice + menu); dragging is not included, since it can be very noisy with many manual movements. Individual participants overlaid as jittered points. Right: within Combined, the same voice/menu split alone - the menu's median is 0 (54\% of participants never touched it), while voice is used by everyone. $N=24$.}
\label{fig:customization-frequency}
\end{figure}

Participants adapted the interfaces most often when conversational control was the only method available. Median adaptation actions per participant were 18.0 in ConverseMR, 11.0 in Combined, and 7.0 in DM (~\autoref{fig:customization-frequency} 6, left). Recall that dragging is not counted, so the DM and Combined values are lower bounds on manual activity, while the ConverseMR value is complete. The ordering therefore understates rather than inflates the gap between speaking and reaching.

Within Combined, participants issued a mean of 9.79 voice commands and 3.21 menu interactions, so voice outnumbered the menu roughly 3:1 and accounted for a mean 81.2\% (median 100\%) of logged activity. The median participant never opened the menu: 13 of 24 (54\%) did not use it once at all once voice was available (~\autoref{fig:customization-frequency}, right), against a median of 7 menu uses when DM was the only method.

\subsubsection{Communication of Adaptation Needs in Natural Language}

We coded each applied command against the ten themes from the formative study (Section~\ref{sec:preliminary-study}), assigning multiple labels where a command expressed several intents and coding unmatched commands as Other.
We used Claude's Sonnet 5 model to support the coding process, as the categories and themes had already been defined in the formative study. Each command was matched against all ten theme definitions based on its parsed action and parameters. Unmatched commands were coded as \emph{Other}. A researcher then reviewed and revised the generated labels to ensure their correctness, and to find out if there is a need for introducing new themes. We found that one frequent pattern had no counterpart in the formative study: \emph{Revision (Replacing a Prior Request)}, where participants explicitly replaced a previous request. We thus introduced a new theme for this. 
The same coding was not possible for DM interactions, where dragging and menu selections do not expose what the user intended specifically. However, the menu and dragging together cover most of the same adaptation space with less flexibility in how an adaptation can be specified.

Table~\ref{tab:themes} reports the share of commands carrying each label. Percentages exceed 100\% because a command could have multiple labels. Anchoring Related rose from 36.8\% to 54\% (67\% in Combined), while the themes that describe behavior more indirectly decreased in frequency (Movement / Gaze Based from 47.4\% to 5\%, Layout Management from 29.8\% to 14\%, and Rule-Based Triggering from 21.1\% to 3\%). In most cases, participants named a target and an anchor (e.g., "Anchor the control panel to my head" (P5), "Put key task UI above box 4" (P8)) rather than describing the dynamic behavior that an anchor produces. World / Ergonomics Knowledge moved in the opposite direction, from 22.8\% to 38\% ("Bring control panel close to me", P4), suggesting that participants were willing to leave placement parameters to the system once they could see and correct the result. Adjustments differ across conditions with 35\% in ConverseMR and only 20\% in Combined where hand dragging was available. Revisions accounted for 27.7\% of commands in ConverseMR and 40.8\% in Combined. For example, P4 first anchored the Key Task UI to their torso, then reanchored it to the head after finding the placement impractical. Such commands could not arise in the one-shot elicitations of the formative study.

\begin{table*}[t]
\caption{Share of applied conversational commands carrying each theme from the formative study. Commands could receive multiple labels, so percentages do not sum to 100\%. \emph{Revision} was not part of the formative taxonomy.}
\label{tab:themes}
\small
\begin{tabular*}{0.8\textwidth}{@{\extracolsep{\fill}}lccc@{}}
\toprule
\textbf{Theme}
& \textbf{Formative}
& \multicolumn{2}{c}{\textbf{Conversational Component}} \\
& \textbf{Rank, \%}
& \textbf{Conversational}
& \textbf{Combined} \\
\midrule
UI Calling                 & \#9, 73.7\%  & 63\%  & 69\% \\
Anchoring Related          & \#3, 36.8\%  & 54\%  & 67\% \\
Reference Points           & \#1, 96.5\%  & 47\%  & 58\% \\
World/Ergonomics Knowledge & \#6, 22.8\%  & 38\%  & 37\% \\
Adjustments                & \#8, 29.8\%  & 35\%  & 20\% \\
Visibility Rules           & \#4, 31.6\%  & 22\%  & 27\% \\
Layout Management          & \#2, 29.8\%  & 14\%  & 11\% \\
Movement/Gaze Based        & \#7, 47.4\%  & 5\%   & 7\% \\
Rule-Based Triggering      & \#5, 21.1\%  & 3\%   & 4\% \\
Deviation Constraint       & \#10, 8.8\%  & 0.5\% & 1\% \\
\midrule
\textit{Revision}          & -           & 27.7\% & 40.8\% \\
\textit{Other}             & -           & 0.8\%  & 0\% \\
\midrule
\textit{Total}             & 57           & 376 & 191 \\
\bottomrule
\end{tabular*}
\end{table*}

\subsubsection{Conversational Interaction Over Time.}

Behavioral logs showed that most voice commands proceeded without clarification, while participants frequently continued adapting the interfaces after a solution was applied. Of the 988 voice commands, 89.6\% passed the confidence check directly, whereas 10.4\% required clarification. Among the 781 commands that reached a concrete solution, the next action split roughly three ways. In 32.4\% of cases, participants \emph{replaced} a property set by the previous command, effectively asking for a new behavior for the same interface (e.g., \emph{``Anchor the Key Task UI to my torso''} followed by \emph{``Anchor it to my head instead''}). In 31.2\%, they \emph{refined} the same interface with an additional adjustment while keeping the previous behavior (e.g., \emph{``Anchor the Control Panel to my hand''} followed by \emph{``Move it a little closer''}). Participants instead moved on to a different interface element in 27.7\% of cases (indicating that they accepted the previous behavior), while 8.7\% repeated the same command on the same target. Beyond these immediate command-to-command transitions, we observed 15 explicit revert events in which participants restored an earlier configuration. Table \ref{tab:command-sequences} summarizes these patterns.
As described in details in Appendix~\ref{app:system-val}, we validated different layers of the system. Speech recognition was the main source of errors, while the LLM-based interpretation resolved most transcription issues through contextual prompting. Overall, participants rated whether the applied commands behaved correctly at a median of 5 out of 7, similar to DM.


\begin{table}[t]
\caption{Command progression and subsequent customization behavior. Percentages for command processing are calculated over all 988 voice commands; percentages for subsequent actions are calculated over the 781 commands that reached a concrete solution. Reverts are reported separately because they represent explicit recovery events rather than immediate post-solution transitions.}
\label{tab:command-sequences}
\small
\begin{tabular}{llrr}
\toprule
\textbf{Stage} & \textbf{Outcome} & \textbf{$n$} & \textbf{\%} \\
\midrule
Command processing
& Passed confidence check & 885 & 89.6 \\
& Required clarification & 103 & 10.4 \\
\midrule
After concrete solution
& Replaced previous property & 253 & 32.4 \\
& Refined same target & 244 & 31.2 \\
& Moved to different target & 216 & 27.7 \\
& Repeated same command & 68 & 8.7 \\
\midrule
Recovery
& Explicit revert & 15 & -- \\
\bottomrule
\end{tabular}
\end{table}

\section{Discussion \& Conclusion}
We set out to understand whether users can control the adaptive behavior of MR interfaces through natural language, and what such control offers over direct manipulation. Participants expressed a broad range of adaptation needs in speech, and used \system{} for most of their adaptations whenever it was available. \system{} alone, however, was never rated highest on any measure. It was statistically indistinguishable from DM on seven of eight usability items and worse on the time spent correcting mistakes, while the combination of both methods was preferred throughout. We discuss below what this pattern means for the design of conversational adaptation, and for the role users can take in adapting MR interfaces.

\subsection{Language Replaces the Menu, Not the Hand}
The Combined condition revealed that participants did not take a choice between conversational control and direct manipulation. Rather, they split their adaptations by the kind of work involved. One kind is deciding how an interface should behave, such as which anchor it should follow, which element a request applies to, and under which condition the behavior should hold. The other is putting an element at a particular spot, ``a little higher'', ``slightly closer'', ``just above the box''.

Language covers the first kind well. Anchors, interfaces, objects in the room and task conditions all have names, so a participant could state a complete behavior in one utterance, as in \textit{``Anchor the control panel to my head''} (P5). Language covers the second kind poorly. A position in space has no name, and the vocabulary available for it is too coarse to carry the value. Participants who wanted an element slightly elsewhere said \textit{``Move key task a little further''} (P6), then looked at the result and said it again. The hand does not have this problem, since the movement is itself the specification~\cite{shneiderman1983}. 

The DM menu offered only the first kind of work. It let participants select an anchor and a coarse distance, which is precisely what they could also say out loud, and saying it required no panel, no search for the right entry, and no selection from drop-down menus. Given both, most participants stopped using it. Thirteen of 24 never opened the menu once voice was available, and anchoring requests rose from 54\% of commands in ConverseMR to 67\% in Combined. Dragging was not replaced in the same way, and \emph{Adjustments} fell from 35\% to 20\% of voice commands once participants could reach for an interface by hand instead.

Choosing between \system{} and DM is therefore note just a choice between coarse and fine adaptations, which is how participants described it, but between what can be expressed through language and what is better shown or gestures through hands. 

\subsection{Designing for Predictable Conversational Adaptation}

\paragraph{Prior experience and failures shape interaction expectations.}
Participants' expectations of ConverseMR were shaped both by their experience during the study and by prior interactions with speech-based assistants. Several spontaneously compared ConverseMR with Siri or Alexa, bringing existing assumptions about what voice interaction can understand and how reliably it should respond. These expectations then changed through use in both directions. Successful commands helped participants predict how to phrase later requests, and some revised an initially negative view once the method paid off: ``At first I thought I didn't like voice... then I saw it's much easier, and I preferred voice'' (P21). A single failure could be enough though to move a participant the other way, as P16 described: ``I lose trust with one failure - I just do it myself [after that].''  A failed interaction therefore affected not only the immediate recovery but the participant's choice of modality for the rest of the session, and their developing model of what the system could do. Early interactions and clear, predictable responses are thus particularly important for helping users learn when conversational adaptation will work and when another method is preferable.


\paragraph{Communication gap between commands and expected outcomes.} Some unexpected outcomes were not caused by incorrect system interpretation or incorrect user input, but by a gap between what a command specified and what the participant expected to happen. For example, P13 asked \system{} to follow them while moving. The system interpreted this as a conditional behavior and followed the participant only while they were in motion (a conditional trigger). While this interpretation was consistent with the command, P13 expected the interface to start following immediately and continue doing so, even when they were standing, like a head-anchored interface. Similar cases occurred when the system applied a command as specified, but the resulting behavior still differed from what the participant had expected. This suggests a communication gap between the literal meaning of a command and the behavior users have in mind. Addressing such cases requires more accurate intent recognition, better ambiguity detection, and a more responsive system that communicates its interpretation clearly, allowing users to identify and correct mismatches during interaction. 

\paragraph{Feedback mechanisms.} Conversational adaptation needs lightweight feedback that makes the system's current state and recent actions visible without interrupting the interaction. While \system{} provided these, participants' suggested improvements  across different stages of interaction: when the system starts listening, when an object is being changed, and when the system is uncertain. For example, P13 preferred an auditory cue for listening because it would be easier to notice than text, and also suggested highlighting or spatialized sound to indicate which object was moving and in which direction. P15 similarly viewed clarification as useful when the system was uncertain. These comments suggest that feedback should not only confirm that a command was received, but also communicate what the system understood and what changed as a result. This becomes particularly important for recovery. If users cannot easily tell what a command changed, deciding whether to undo and knowing what will be undone also becomes difficult. A lightweight history of recent adaptations, combined with clear multimodal confirmation and an explicit undo action, could therefore make conversational adaptation more predictable and easier to recover from without requiring users to ask the system for its current state.

\subsection{Toward User-Led MR Adaptation and Beyond}
Our log data shows that participants did not simply state an adaptation intent, receive a solution, and stop. Requests sometimes needed clarification before a solution could be formed, and applied solutions were then refined, replaced, repeated, or reverted. 
Participants only produced these once they could see an applied change, which suggests that the loop closes not at the command but at the adaptation the user observes.
We read this as evidence that adaptation of MR interfaces need not be either
something a system does to a user or something a user configures once. Users can instead state what they want the interface to do and leave the system to work out how, revising the policy as their situation changes. User-led adaptation thus sits between automatic adaptation and manual customization. It is an iterative process between a user and a system, aligning the user's intent with the system's solution. Clarification, refinement, replacement, and recovery are therefore not exceptional error cases but central parts of that loop, which means designing for user-led adaptation is designing for the loop rather than for the command. The same structure may apply beyond MR adaptation to tasks such as scheduling, trip planning, resource allocation, or personal configuration, where users begin with an incomplete or implicit intent and progressively shape a solution provided by the system~\cite{Zhang2026}. Our transition categories offer a first empirical vocabulary for describing and evaluating such intent-driven systems, with \system{} as one concrete realization.

This iterative behavior also gives customization frequency two possible interpretations. More actions can be positive, indicating that participants used the available flexibility to make more adaptations, but they can also reflect difficulty reaching the intended result, requiring users to refine or replace earlier changes. This distinction is visible across conditions. \system{}-only had the highest frequency ($Mdn=18.5$), but also lower subjective ratings and more replacements, suggesting that at least some of these additional actions reflected correction or refinement. Combined had a lower frequency ($Mdn=11.0$) despite offering greater flexibility, suggesting that participants could make complex adaptations with fewer steps. DM had the lowest frequency ($Mdn=7.0$), which may instead reflect its more constrained adaptation space. Thus, higher customization frequency is not inherently better or worse, but needs to be interpreted alongside how and why participants continued adapting.


\subsection{Limitations and Future Work}
Our study lasted only 90 minutes, which means our results describe early rather than habituated use. Participant's comments showed that they were still learning what ConverseMR could understand. In some cases, a single failure was enough to move some of them to the manual alternative for the rest of the session. Since Combined always came third in our study design, its high ratings might be partially conflated with longer experience. Long-term studies are needed to explore how the use of conversational systems changes users' preferences and interactive behavior as people gain more experience with them. 

The study room was compact ($5 \times 3.5$\,m), so walking to an interface cost participants little. ConverseMR's reach advantage is likely understated here and might grow in larger spaces or mobile outdoor conditions. The same holds for the environments themselves. Our scene was sparse and its objects were labelled to make spatial references easy, whereas real spaces offer many more anchors and no labels, and resolving a reference such as ``next to the shelf'' becomes harder as the number of candidates grows. Further studies in larger more varied settings might reveal different considerations and preferences. 

The DM baseline is one of several defensible choices. We combined the menu with manual spatial movement to give participants the full range of manual control, but the two could have been separated into conditions of their own, and the menu-only and dragging-only ratings in Section~\ref{sec:results-quant} suggest they are experienced differently.

The adaptation space we supported with either method also bounds what participants could ask for. In some cases, participants requested anchors we did not offer, in particular objects and surfaces in the environment and further locations on the body, as well as layouts they could save and recall. Several suggested gesture shortcuts for frequent adaptations such as anchoring, which would sit between speaking and navigating a menu. A system that recognises which adaptations a user requests in which situation could offer them again when the situation recurs, which would spare users from restating the same policy and would still leave them in charge of what the interface should do. 

Beyond these extensions, our results point to three important directions for future work: handling domain-specific terminology and  ambiguity in language, making both the scope and the outcome of an adaptation legible in the interface, and closing the latency gap between stating an adaptation and seeing it applied, which participants experienced as a frustration distinct from being misunderstood.

\bibliographystyle{ACM-Reference-Format}
\bibliography{references}

\appendix

\section{Questionnaire Items and Provenance}
\label{app:questionnaire}

Table~\ref{tab:items} lists every closed-ended Post-Method and Final item and its provenance. Items marked \emph{adapted} follow the cited instrument's wording closely; items marked \emph{study-specific} have no external validated source and are justified by the cited construct/framework instead.

\begin{table*}[t]
\caption{Closed-ended questionnaire items, their exact wording, and their provenance.}
\label{tab:items}
\small
\begin{tabular*}{\textwidth}{@{\extracolsep{\fill}}p{0.16\linewidth}p{0.42\linewidth}p{0.26\linewidth}l@{}}
\toprule
\textbf{Item} & \textbf{Question wording} & \textbf{Provenance} & \textbf{Source} \\
\midrule
\multicolumn{4}{l}{\textit{Post-Method questionnaire (administered after each block; Table~\ref{tab:likert})}} \\
\midrule
Met requirements & ``This method's capabilities met my adaptation requirements.'' & adapted (UMUX) & \cite{finstad2010umux} \\
Easy to adapt & ``Overall, making adaptations with this method was easy.'' & adapted (UMUX); cf.\ SEQ & \cite{finstad2010umux,sauro2009seq} \\
Comfortable & ``This method was comfortable to use.'' & study-specific (ergonomics) & \cite{belo2021xrgonomics} \\
Felt in control & ``I felt in control of how the interfaces was placed and behaved.'' & study-specific (DM theory) & \cite{shneiderman1983direct} \\
Behaved as expected & ``This method behaved as I expected.'' & study-specific & \cite{walkerparadise} \\
Fast enough & ``I could make adaptations quickly enough with this method.'' & study-specific; cf.\ temporal demand & \cite{hart1988nasatlx} \\
Time spent correcting & ``I had to spend too much time correcting things with this method.'' & adapted (UMUX) & \cite{finstad2010umux} \\
Supported wanted adaptations & ``This method supported the kinds of adaptations I wanted to make.'' & study-specific (adaptation objectives) & \cite{belo2022auit} \\
\midrule
\multicolumn{4}{l}{\textit{Final questionnaire (administered once, after all three blocks)}} \\
\midrule
Overall want to use each method & ``Overall, how much would you like to use each method?'' (rated separately for Conversational, DM, Combined) & study-specific & - \\
Preferred method per interface element & ``For each interface, which method did you prefer?'' (rated separately for Key Task UI, Visual Task UI, Control Panel) & study-specific (task design) & - \\
Simple-placement bipolar item & ``Which method was better for simple one-step placement changes?'' & study-specific (DM theory) & \cite{shneiderman1983direct} \\
Complex-adaptation bipolar item & ``Which method was better for complex adaptations?'' & study-specific (optimization-based adaptation) & \cite{belo2022auit} \\
\bottomrule
\end{tabular*}
\end{table*}

\section{System Validation}
\label{app:system-val}

Across the full corpus, we examined accuracy along 
the intent $\to$ transcript $\to$ structured representation $\to$ executed adaptation $\to$ perceived outcome pipeline.
For speech recognition, we aligned logged transcripts with the notes taken by the experimenter on what participants actually said as the ground truth utterances. We found that of these, only 49.1\% were transcribed correctly by the Meta SDK, 37.4\% were partially preserved, and 13.5\% were totally incorrect. After the LLM-based interpretation stage, most of this inaccuracy could be resolved though. For high-confidence commands that did not require clarification, the objective extracted into the intermediate representation matched the ground truth objective in 96.0\% of cases, suggesting that the second stage of the pipeline could recover from many speech-recognition errors introduced in the first stage.
It was consistent with high confidence ($M=0.845$, $Mdn=0.86$) and infrequent clarification (11.2\% of 916 decisions). Execution was similarly stable: only 13.4\% of 543 consecutive applied-command pairs immediately replaced a property just set on the same target, an upper bound that also includes normal refinement, and only 15 explicit reverts occurred. 
 
Finally, participants rated whether the system behaved as expected at $Mdn=5.0/7$, statistically indistinguishable from DM (5.5) and below Combined (6.0). Overall, most observed errors arose during speech recognition, while the intent extraction stage was tolerant to such problems, and it could resolve the noisy speech recognition transcripts. Particularly for study-specific terminology, while interpretation and execution were considerably more reliable once usable input reached the system.

\section{Intermediate Representation}
\label{app:ir}

To translate conversational commands into adaptations that can be executed by \system{}, we use an intermediate representation between the LLM interpretation and the adaptation engine. For each command, the LLM maps the ASR transcript to a single JSON object conforming to a fixed schema, enforced through the OpenAI Responses API using strict structured output. The representation specifies the targeted interface element(s), adaptation objective(s), spatial and behavioral constraints, and control information such as clarification, conflict resolution, and conditional triggers. The runtime operates only on this structured representation and does not directly parse the raw ASR transcript for adaptation logic. Fields that are not applicable to a command are assigned neutral defaults.

\paragraph{Supported objectives.}
The intermediate representation supports the following adaptation objectives:
\begin{itemize}
\item \textbf{Visibility and reachability:} Make Visible, Make Reachable, Field of View, Look Towards, and Constant View Size.
\item \textbf{Spatial constraints:} Fix Overlap, Avoid Occlusion, Avoid Collision, Distance Interval, Spatial Coherence, and Surface Magnetism.
\item \textbf{General adaptation:} Adapt, Optimize, Reset, and Undo.
\item \textbf{Context-aware adaptation:} Avoid Adaptation While Moving.
\end{itemize}

\subsection{Example Representations}

The following examples illustrate how conversational commands are mapped to the structured representation. The first shows a simple spatial relationship between an interface and a physical reference, while the second demonstrates a compositional command that specifies different adaptations for two interfaces.

\begin{lstlisting}[
caption={Example A: place Visual Task UI above Box~3.}]
{
"raw": "put the visual task on top of box three",
"domain": "auit",
"target": "visual",
"targets": [],
"app_action": "apply",
"clarify_options": [],
"conflict_resolution": "merge_additive",
"conflicts_with_prior": false,
"trigger_condition": "none",
"clear_triggers": false,
"user_message": "Visual Task UI placed above Box 3.",
"adaptation_summary": "place visual above box_3",
"response_text": "Done.",
"confidence": 0.9,
"action": "",
"priority": "",
"params": {
"ui_intent": "adapt",
"ui_intents": [],
"anchor_target": "unchanged",
"visibility": "unchanged",
"keep_in_view": "unchanged",
"lock_world_pose": "unchanged",
"unlock_pose": "unchanged",
"min_distance_meters": 0,
"max_distance_meters": 0,
"keep_in_view_angle_deg": 0,
"reference_label": "box_3",
"secondary_reference_label": "unchanged",
"spatial_relation": "above",
"body_side": "unchanged",
"constraints_to_clear": [],
"move": {
"mode": "none",
"world_position": { "x": 0, "y": 0, "z": 0 },
"delta_meters": { "x": 0, "y": 0, "z": 0 }
}
},
"additional_commands": []
}
\end{lstlisting}

\begin{lstlisting}[
caption={Example B: compositional command -Key Task UI remains in view while Control Task UI follows the torso.}]
{
"raw": "key task in front of me and control panel follow me",
"domain": "auit",
"target": "key",
"targets": [],
"app_action": "apply",
"clarify_options": [],
"conflict_resolution": "merge_additive",
"conflicts_with_prior": false,
"trigger_condition": "none",
"clear_triggers": false,
"user_message": "Key Task UI in view; Control Task UI follows you.",
"adaptation_summary": "FOV key; anchor control torso",
"response_text": "Done.",
"confidence": 0.85,
"action": "",
"priority": "",
"params": {
"ui_intent": "field_of_view",
"ui_intents": [
"field_of_view",
"make_visible",
"look_towards"
],
"anchor_target": "unchanged",
"visibility": "visible",
"keep_in_view": "true",
"lock_world_pose": "unchanged",
"unlock_pose": "unchanged",
"min_distance_meters": 0,
"max_distance_meters": 0,
"keep_in_view_angle_deg": 0,
"reference_label": "unchanged",
"secondary_reference_label": "unchanged",
"spatial_relation": "unchanged",
"body_side": "unchanged",
"constraints_to_clear": [],
"move": {
"mode": "none",
"world_position": { "x": 0, "y": 0, "z": 0 },
"delta_meters": { "x": 0, "y": 0, "z": 0 }
}
},
"additional_commands": [
{
"target": "control",
"ui_intent": "",
"ui_intents": [],
"anchor_target": "torso",
"spatial_relation": "unchanged",
"body_side": "unchanged",
"keep_in_view": "unchanged",
"min_distance_meters": 0,
"max_distance_meters": 0,
"move": {
"mode": "none",
"world_position": { "x": 0, "y": 0, "z": 0 },
"delta_meters": { "x": 0, "y": 0, "z": 0 }
}
}
]
}
\end{lstlisting}










\end{document}